# A System Framework for Smart Class System to Boost Education and Management

Ahmad Tasnim Siddiqui
Department of Computer Science,
Taif University, Taif
Kingdom of Saudi Arabia
Zip – 21974

Dr. Mehedi Masud
Department of Computer Science,
Taif University, Taif
Kingdom of Saudi Arabia
Zip – 21974

*Abstract*—The large number of reasonably priced computers, Internet broadband connectivity and rich education content has created a global phenomenon by which information and communication technology (ICT) has used to remodel education. E-learning can be explained as the use of available information, computational and communication technologies to assist learning practice. In the modern world, education has become more universal, and people are looking for learning with simplicity and interest. Students are looking for more interactive and attractive learning style rather than old traditional style. Using technological learning, we can enhance the education system. We can deliver quality education to students as well as we can ease and uniform the process of education by using the modern technologies and methods. In this paper, we propose a smart class model to manage the entire educational activities and hence to enhance the quality of education.

## I. INTRODUCTION

This paper describes the processes of developing and implementing the framework of Smart Class system using e-learning technologies for learning. The main purpose is to raise the competence of quality education of the society for lifelong learning [15]. In this competitive and globalized world, there exists a tremendous focus on having e-learning and e-technology in a variety of working sectors. However, you can find still much scope for improvement on E-learning competency among the scholars, specifically those from rural areas who sadly are still struggling to get the knowledge. Effective work is urgently needed in our educational system in order to secure the futures of students, so that they can remain competitive in the job market. According to the Merriam-Webster dictionary, "Learning is knowledge or skills acquired by instruction, study or experience [1]". We can also explain the above line as: A continuous process of acquiring knowledge and improving our skills either by practice, experience, study or by being taught by somebody. Using e-learning, we can provide the quality education to remote and rural regions with the help of modern technologies like satellite, internet, and mobiles [2]. Concentration is very much requiring for learning anything. It is a fact; current students are vastly distinct from how they were a long time ago. Reported by Eaton [6], today's students are very much tech-savvy. They're able to access a whole lot of resources and information just at their finger tips. They are hungry for motivation, inspiration, and guidance.

E-learning involves a very wide range of applications. It includes computational, communication technologies along with other modern devices like interactive TV etc. [2] Smart class system is entirely different from the traditional way of teaching by writing on black boards. It is a modular approach specially designed to help faculties, instructors to compete with new challenges and developing students' capabilities and performance [3]. Smart class can be defined as the improved way of education in which teachers teach and students learn in colleges or universities with advanced and significant use of technology. Smart class means to use the technology right in the way for faculties or instructors in the classrooms or in the laboratories. Students are able to learn and understand difficult concepts and understand the complex problems by watching highly effective audio visuals and animations. By using these we can make learning a fun for the students which will definitely improve their overall performance. Smart class system also enables faculties or instructors to rapidly evaluate the learning by their students in class. The system can automatically mark attendance of students, faculties and instructors by just swapping the smart card and many other activities. Smart class systems are also supposed to be environmental friendly, so that they can provide good environmental practice for the students as well as faculties and instructors.

### A. Features of Smart Class System

Smart Class system is a key solution which is intended to support faculties and teaching assistants to overcome with their daily classroom and lab challenges and also improving student's academic interest and performance with easy, practical and significant use of technology. Smart Class helps faculties to make sure that every student in the class is getting knowledge, by providing the wide range of learning patterns in the classroom and in lab sessions. It is also very helpful in managing student's interest and engagement in learning within the classroom. Smart Class makes the problems easy for teacher, abstract curriculum concepts which are difficult to understand and imagine for students or relate by the use of 3-D (three dimensional), interactive multi-media approach.

Smart class have many benefits like: i. Faculty/instructor spends more time in teaching rather than time consuming in getting started. ii. Can share all the forms, tests, quizzes and assignments to students in just a click. iii. System can collect files automatically after students are finished. iv. Last Class View & Planner, Class Regulation& Monitoring, Assessment of the work, Share Student Screen and Sharing Faculty/Instructor Screen. The term significant use of technology involves a very wide range of technological resources. It includes computers, smart phones, tablets, Internet and web sites, virtual classrooms, projectors, smart boards (interactive white boards, touch screen LCD) and digital association. By providing technological and quality educations along with the quizzes and some other interesting activities, we can provoke the students to learn easily. Smart class also provides a consistent messaging. It minimizes the problems related with different instructors teaching style. Because everyone has their own teaching style and different knowledge base on the same subject. Smart class approach can provide tremendous improved output.

## II. LITERATURE REVIEW

E-learning is recognized worldwide in the form of easy learning approach. This is the learning procedure which is delivered through internet, laptops and wireless mobile handheld devices which allows learning anytime and anywhere. Electronic learning takes learning to persons, communities and countries have got previously too remote, socially or geographically, for other categories of educational initiative [7]. E-learning is usually thought as the usage of available information and communication technologies to facilitate learning process. E-learning is a combination of learning as well as the Internet technology [8].

With the advancement in learning technology and change in school education system, it is essential for educators to analyze the classroom situation all the way through the lens of digital interaction. They should propose integrated teaching solutions that put tools such as smart screen, document sharing, on the spot polling and surveys, and remote device management directly by the hands of students and faculties [4].

Electronic learning can be viewed not just being a silver bullet in education, but more as something for learners to gain access to knowledge while there're on the go at anytime and anyplace with full flexibility. Smart class room enabled with interactive white board supports non-linear learning in two different ways: (1) It provides accessing of hypertext and hypermedia online or as external files and (2) It can be accessed by moving back and forth for review slides related to questions/answers of the students and faculties [16]. A lot more educational facilities of the universe are usually on your journey to e-learning and mobile learning so as to take advantage of the ubiquitous quality could possibly offer for educational purposes. An emerging body of literature that explores possibly mobile learning for educational contexts has identified several significant features about mobile learning for example convenience, cost effectiveness, motivation to know, flexibility, accessibility and also the interaction [9][10][11][12].

Figure 1 explains the U.S. example of growth in tablet, laptops and e-reader expenditure over time. These devices, which make up only 7% of PC spending in the U.S. in 2011, will be increased around 24% of spending by the end of 2016. The increase is mostly due to the advancement in educational technology.

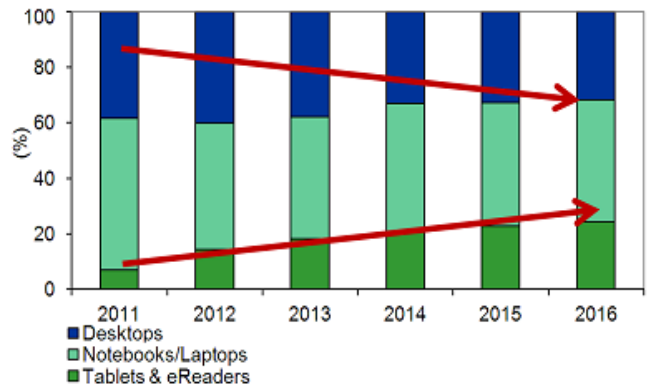

Fig. 1. U.S. Education system spending on Laptops and Tablets as a Percent of Total Spending, 2011-2016 (Source: IDC Global Technology and Research Organization, 2012)

Learner's ability to grasp the knowledge depends on the method of teaching. Figure 2, learning pyramid explains learning retention rates by type of teaching. These days the students have grown up enough with the Internet which provides immediate access to a wealth of information and in multiple formats available such as audio, video, images and text [19].

In the given figure 2, we can see that retention rate is very high in participatory learning as compared to the passive learning which less interactive or non-interactive. Learning through lecture only is very less which is only 5%, while average retention rate is through group discussion which is 50% while learning with practice is 75% and learning with the immediate use of learning that means the applied learning retention rate is 90%. Here the percentage shows the average grasping rate of students.

According to Knight et al., we have to differentiate between the linear and traditional interactivity of teaching and learning where faculty and students interactions as well as interactions of students with their peers takes place and technical interactivity which involves physical interaction with the electronic devices e.g. laptops, tablets, e-pads, e-readers and Smartphone etc [18].

Technical expertise and interactivity can encourage the practice of skills, while pedagogical interactivity provides ability to higher-order thinking and reflections on the learning process [17].

But according to Cisco, these oft-quoted statistics are unsubstantiated [5]. Below is the Cone of Learning which is developed by Bruce Hyland from material by Edgar Dale.

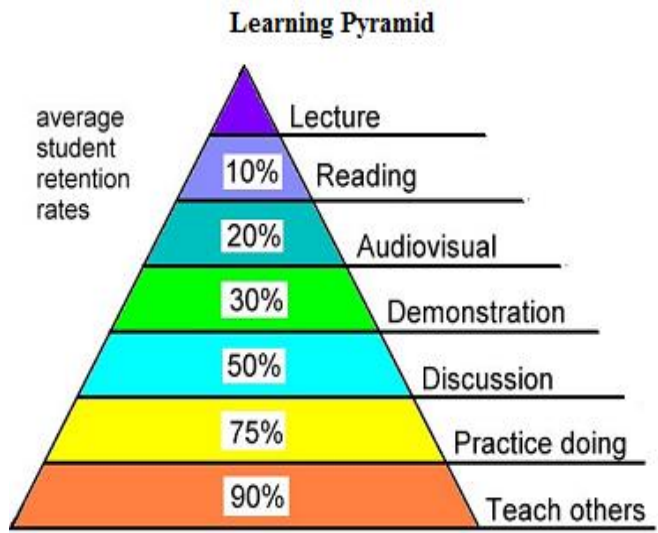

Fig. 2. Learning Retention Rates by Type of Teaching (Source: The World Bank and the National Training Laboratories, Bethel, Maine)

But From the figure 2, one thing is clearly indicated that learning which is interactive and immediately used is more effective than the traditional learning. So, we can make learning more interactive by using advanced technologies.

### III. OUR CONTRIBUTION

The main aim of this research is to present a system model for enhancement of education level as well as interest in education. The priority is to provide interactive, uniform academic style and high quality in education. Apart from the traditional black board writing education system we propose a real enhanced education system which is smart class system. It uses more and more advanced technologies to help and support to students as well as faculties and instructors. It uses some inter active learning like forums; quizzes etc. to make the education more interesting, to help students for improving their knowledge level and interest. Students can get proper guidance with the faculties. They can understand and learn every problem easy or complex by audio visual effects, animations and 3-D models. They can get benefits from the special lectures on specific topics by the subject experts. This paper presents a smart class system to enhance the education. Smart classroom covers the whole course of the University/Boards which is flawlessly integrated into the syllabus. The resources include broad learning modules, educational competitions, and interactive e-books, audio, video and soft copies of the study materials in the form of ".pdf" and ".doc(x)" files. In this model we propose smart student tracking system. Every student can be recognized and tracked by the smart card or facial recognition. They can be marked as present when they enter into the class rooms.

The system can also send email and SMS to the parents if students are coming to the classes or bunking schools and colleges. The system also has provision of sending progress reports to the students as well as their guardians.

### IV. PROPOSED E-LEARNING SYSTEM ARCHITECTURE

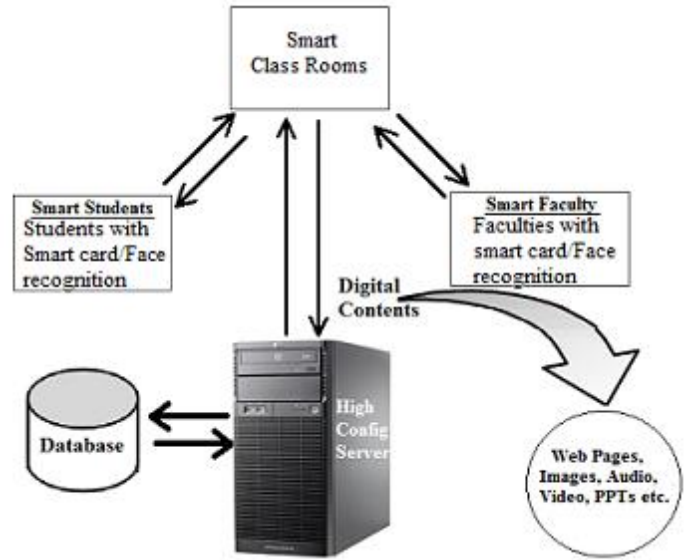

Fig. 3. A Model Framework for Smart Class System

The figure 3 explains our proposed e-learning system; there are four important components in the system. Apart from these four there are two sub components which are used for query processing and sending emails and messages to the students as well as their parents. They are smart class, faculties with smart cards (smart faculties) and students with smart card (smart students). Apart from these main components system includes a database and high configuration server. Each module is integrated. Smart student and smart faculty are directly connected to smart class module.

- **Smart Class:** The first component is Smart Class Room which contains a normal class room equipped with all the modern technologies like computers/laptops, smart phones, interactive TV's, card readers, facial recognitions, high resolution video cameras etc. Whenever any faculty, instructor or student enters into the class he/she has to show the smart card to the card reader or he/she can be recognized by the facial recognition system which marks the attendance of that particular faculty, instructor or student for that particular session. Smart class room is always connected to a high configuration server and a database. Server provides on demand contents to the class room faculties and instructors. The contents are in the form of images, pdf files, 3-D images and models, word document files, audio and video etc. Through the system faculties and instructors can interact to students to explain the topic, to clear the doubts and for other problems. In smart class the faculty or instructor can logged into their session and start the lectures. No need to carry attendance sheets, lecture notes, white board marker and other stuffs which they carry into the normal class rooms. They can browse for the subjects and topics and start the session. They can explain the topics on interactive TV's or monitors. The basic database tables are smart class, students and faculty.

- **Smart Faculty:** The second component is Smart Faculty. It means Faculties with Smart Cards. Smart faculties can logged into the respective session and can perform the required tasks like attendance management, lectures, notes, quizzes and other important activities. They can interact to students to explain the topics, to solve their problems in smart class rooms. They can also see the questions posted by the students into their respective subjects. After answering them they can submit and email provider can send email instantly to the students and status of the question is changed from pending to answer.

- **Smart Student:** The third component of this model is Smart Students. Smart students contains a smart card by which they are able to enter into the class, they are also eligible to mark for attendance by showing the smart card inside the smart class rooms. Smart students can always be connected to the smart faculty by using smart class features. They can understand the problems, they can ask questions and they can also post the questions to the concern smart faculty.

- **Content Centre:** The Content Centre module stores the whole educational materials in to the content database which usually are multimedia contents like texts, images, audios, and videos. Materials can be retrieved online through web portal using some secure web services. After this step the contents are delivered to the learners based on their learning style group.

- **Query Processing**: This is a student based application which enables students to have instant feedback and help from the lecturer with regards to the subjects.

- **Alert Messages**: This is the module, in case of instant communications needed, responsible for sending alert messages using SMS and email to the students as well as their parents.

- **Real-time Interaction:** The model represents the ability to support the teaching interaction and human-computer interaction of the Smart class. This involves handy operation and smooth interaction. In smooth interaction, the Smart class should fulfill the interactive needs of the multi-terminal, and a large amount of data. In interactive session, smart class has the feature to record and store the basic data among faculties' students and computer, to support the teachers and students' self-assessment.

These modules worked together and connected to the database [14]. Entire application is hosted at high configuration server, which is able to serve many classrooms, faculties and students at the same time. Through this application every student and staff member can access the digital contents including web pages, images, audio, video, power point presentations etc. after successful login. They can access the content within the smart class rooms.

## V. CONCLUSIONS

In this paper we have presented a model framework for Smart Class Room. Smart Class Room is designed to help faculties, instructors to compete with new challenges and developing students' capabilities and performance. Institutions and organizations financial condition will force them to contemplate adopting a simple and inexpensive solution. This model provides improved way of education in which teachers teach and students learn in colleges or universities with advanced and significant use of technology. They can interact directly without any hesitations. Smart class has many benefits to the students and faculties. It is very clear that innovation in technology is impacting everywhere and bringing new opportunities for schools, universities and educationalists. The system architecture proposed differs in such a way that it provides flexible learning style which is adaptable to students' favorite learning styles. It means to offer a personalized learning environment which suits individual's learning style. Smart class system helps to increase the learning abilities. It can also be used as a alternative learning method to teach the different IQ level students. There must be technological strategy for the classes, schools and entire learning atmosphere. We can help students as well educators by using advanced technologies to make the future bright.